\documentclass[FBSedit,FBSmath,ecsub]{FBSsuppl}
\usepackage{amsfonts}
\usepackage{amssymb}
\usepackage{epsfig}
\def\qed{\hbox{${\vcenter{\vbox{			
   \hrule height 0.4pt\hbox{\vrule width 0.4pt height 6pt
   \kern5pt\vrule width 0.4pt}\hrule height 0.4pt}}}$}}

\def\Journal#1#2#3#4{{#1} {\bf #2}, #3 (#4)}

\def\LNC{\em Lett. Nuovo Cim.}

\def\NPA{{\em Nucl. Phys.} A}

\def\PRC{{\em Phys. Rev.} C}

\def\ZPA{{\em Z. Phys.} A}

\def\EPJA{{\em Eur. Phys. J.} A}

\title{\hfill{\small {\bf MKPH-T-02-12}}\\
Final State Interaction Effects in Incoherent Pion Photoproduction
on the Deuteron
\thanks{Supported by Deutsche Forschungsgemeinschaft (SFB 443).}
} 
\author{E.M.\ Darwish\thanks{Supported 
by Deutscher Akademischer Austauschdienst. \textit{Present 
address:} Physics Department, Faculty of Science, South Valley 
University, Sohag, Egypt.}, H.\ Arenh\"ovel, M.\ Schwamb}
\institute{Institut f\"ur Kernphysik, 
Johannes Gutenberg-Universit\"at, 55099 Mainz, Germany}

\begin{document}

\maketitle
\vspace*{-.5cm}

\begin{figure}[h]
\vspace*{-.2cm}
\centerline{\epsfig{file=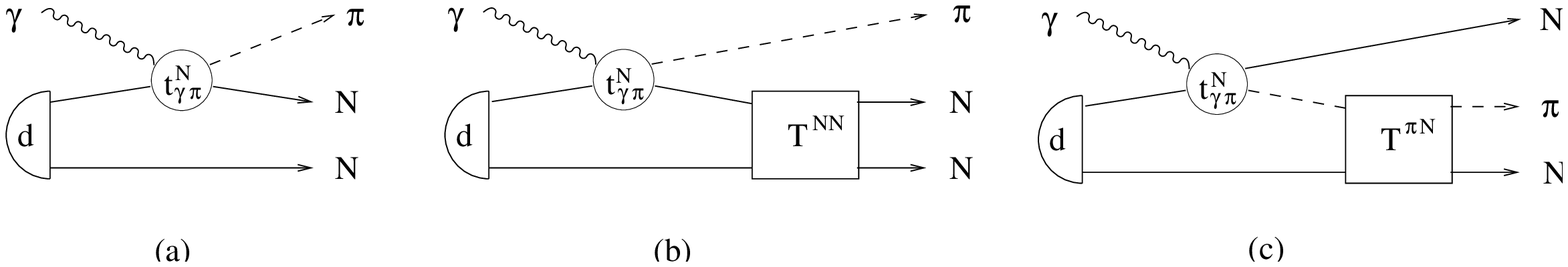,width=7cm}}
\vspace*{-.3cm}
\caption{Diagrams for $d(\gamma,\pi)NN$: (a) IA, (b) $NN$
rescattering, (c) $\pi N$ rescattering.}
\label{fig1}
\centerline{\epsfig{file=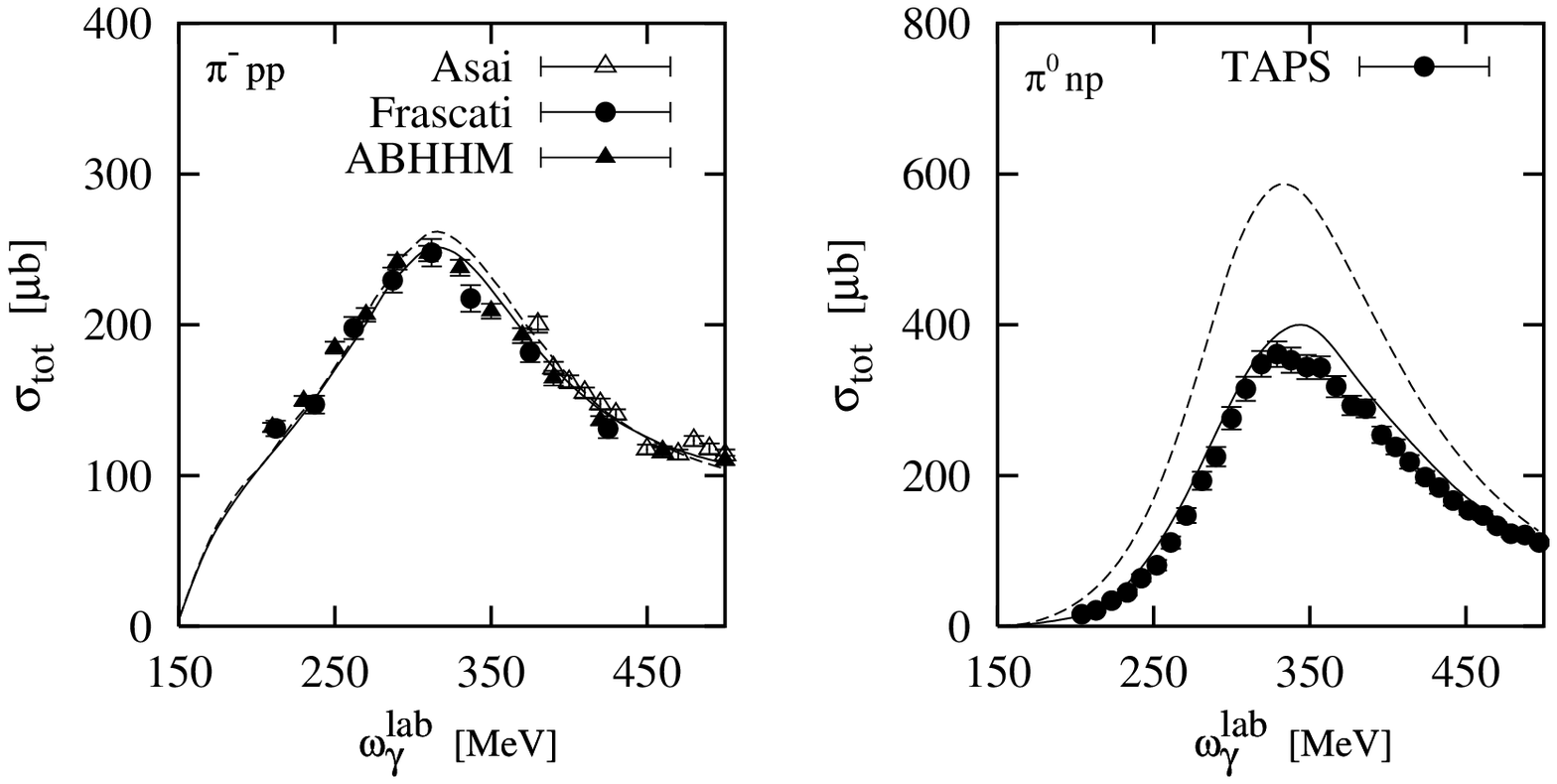,width=7cm,angle=0}}
\vspace*{-.4cm}
\caption{Total cross section for $d(\gamma,\pi)NN$. 
Dashed: IA; solid: IA+complete rescattering; data: Asai 
{\it et al.}, \Journal{\PRC}{42}{837}{1990}, Benz {\it et al.}, 
\Journal{\NPA}{65}{158}{1973} (ABHHM), Chiefari {\it et al.},  
\Journal{\LNC}{13}{129}{1975} (Frascati).}
\label{fig2}
\vspace*{-.3cm}
\end{figure}

\begin{figure}[h] 
\vspace*{-.2cm}
\centerline{\epsfig{file=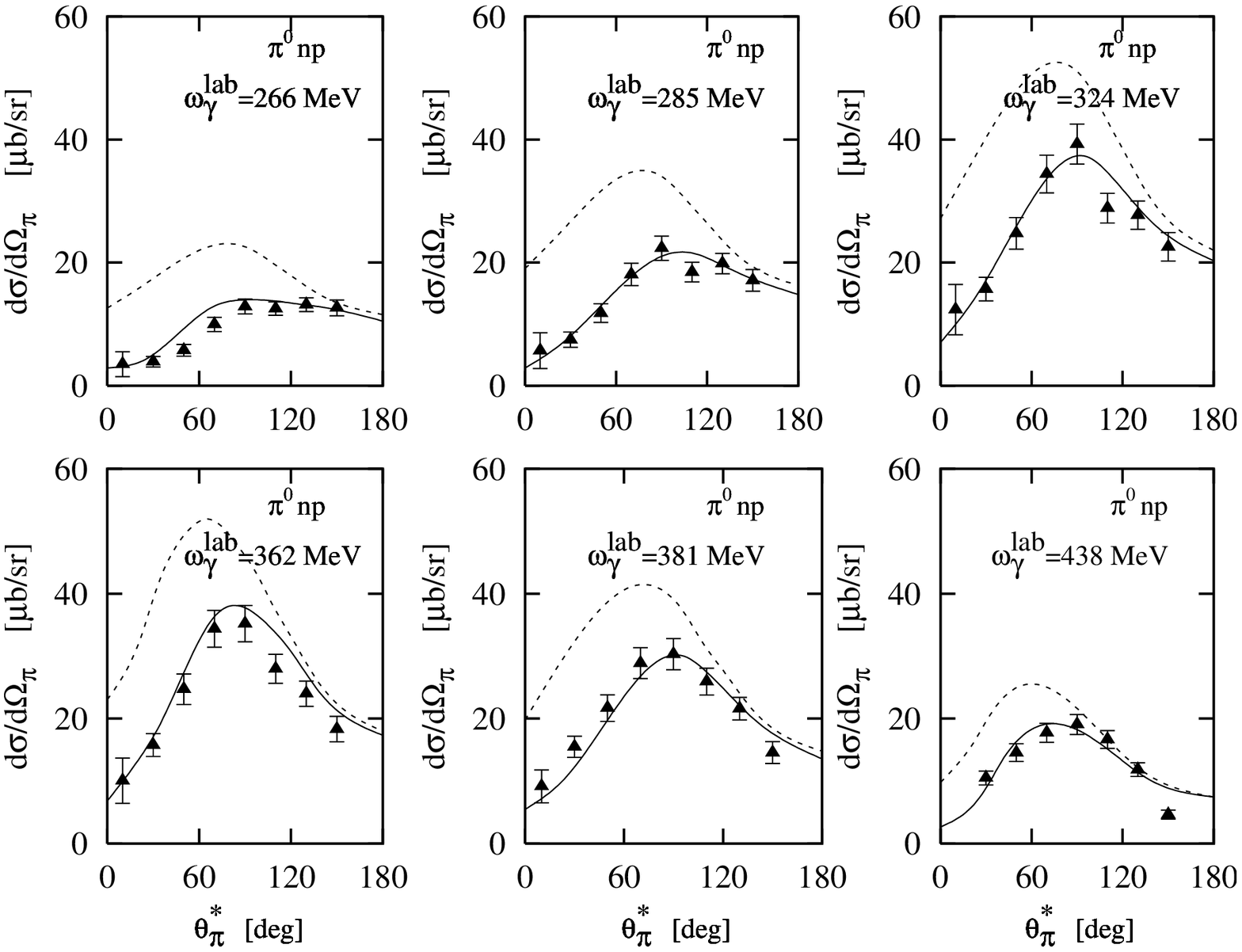,width=7cm}} 
\vspace*{-.4cm}
\caption{Differential cross sections for $d(\gamma,\pi^0)pn$: 
Dashed: IA; solid: IA+complete rescattering; 
Exp.: Krusche {\it et al.}, \Journal{\EPJA}{6}{309}{1999}.}
\label{fig3}
\centerline{\epsfig{file=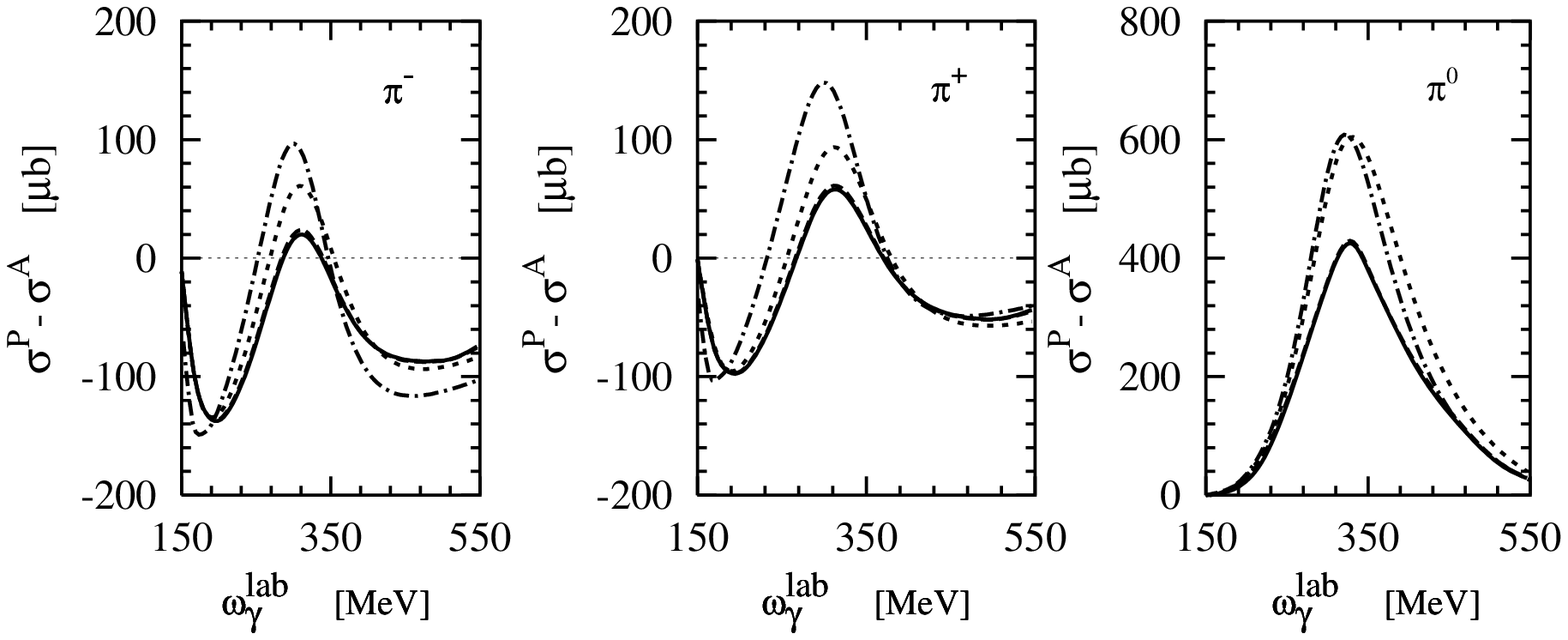, width=10cm}}
\vspace*{-.5cm}
\caption{Spin asymmetry of $d(\gamma,\pi^0)pn$: Dotted: IA; dashed:
IA+NN rescattering; solid: IA+NN+$\pi N$ rescattering. Dash-dotted: 
results for the nucleon, for $\pi^0$ average of proton and neutron 
(right panel).}
\label{fig4}
\vspace*{-.5cm}
\end{figure}

In the present work (nucl-th/0208030; {\it Eur.\ Phys.\ J.} A (in print)), 
incoherent photoproduction of pions on the deuteron in the $\Delta$(1232)
resonance region is investigated where besides the impulse
approximation (IA) complete two-body rescattering in the
nucleon-nucleon ($NN$) and pion-nucleon ($\pi N$) final state
subsystems is included (see Fig.~\ref{fig1}). The elementary $\gamma
N\rightarrow \pi N$ amplitude including Born and $\Delta(1232)$
resonance contributions is taken from previous work of R.\ Schmidt
{\it et al.} (\Journal{\ZPA}{355}{421}{1996}). 
The $t$-matrices for $NN$- and $\pi N$-scattering are obtained from the 
separable 
interactions of Haidenbauer {\it et al.}
(\Journal{\PRC}{30}{1822}{1984}) and Nozawa {\it et al.}
(\Journal{\NPA}{513}{459}{1990}), 
respectively. Final state interaction effects are found to be significant
in total (see Fig.~\ref{fig2}) and differential cross sections (see
Fig.~\ref{fig3}), in particular for $\pi^0$ production, and lead to a
satisfactory agreement with experimental data. 
The large effect for $\pi^0$ production stems mainly from
the fact that in IA quite a fraction of the coherent production is
included due to the non-orthogonality of the final plane wave to the
bound state. 
We have also evaluated the contribution to the spin
asymmetry of the total cross section which determines the GDH-sum rule. 
While in IA the spin asymmetry for $\pi^0$ production 
on the deuteron is very similar to the average over neutron and proton 
asymmetries this is not the case for charged $\pi$ production due to the 
Pauli principle. Moreover, final state rescattering is quite important and 
cannot be neglected (Fig.~\ref{fig4}). For charged $\pi$ production
the effect is much larger in the asymmetry than in the total cross
section and comparable to the effect in the neutral $\pi$ production. 
For the GDH integral up to 550 MeV, we obtain from $d(\gamma,\pi)NN$
in IA $I_{d(\gamma,\pi)NN}^{(IA)}=187~\mu$b which is reduced to
$I_{d(\gamma,\pi)NN}^{(IA+NN+\pi N)}=87~\mu$b by final state
rescattering. 


\end{document}